\documentclass{jpconf}

\usepackage[dvipdfmx]{graphicx}
\usepackage{bm}
\usepackage{amsmath}
\usepackage{amssymb}
\usepackage{mathtools}
\usepackage{tikz}
\usepackage{caption}
\usepackage{subcaption}
\usepackage[utf8]{inputenc}
\usepackage{color}
\usepackage{hyperref}
\usepackage{amsfonts}
\usepackage[subnum]{cases}
\usepackage{ulem}

\newcommand{\newc}{\newcommand}

\newc{\dd}{\text{d}}
\newc{\pa}{\partial}
\newc{\ba}{\begin{eqnarray}}
\newc{\ea}{\end{eqnarray}}
\newc{\red}[1]{{\color{red}{#1}}}
\newc{\blue}[1]{{\color{blue}{#1}}}
\newc{\no}{\nonumber}
\newc{\lt}{\left}
\newc{\rt}{\right}
\newc{\vm}{v_\text{max}}
\newc{\vmin}{v_\text{min}}
\newc{\om}{\omega}
\newc{\Om}{\Omega}
\newc{\mph}{m_\Phi}
\newc{\Mph}{M_\Phi}

\begin{document}

%\preprint{}
\title{Greybody Factor for massive scalar field in charged black hole
}

\author{
Ratchaphat Nakarachinda
}

\address{The Institute for Fundamental Study (IF), Naresuan University, 99 Moo 9, Tah Poe, Mueang Phitsanulok, Phitsanulok, 65000, Thailand}

\address{Department of Mathematics and Computer Science, Faculty of Science, Chulalongkorn University, 254 Phaya Thai Rd., Pathum Wan, Bangkok, 10330, Thailand}

\author{
Suppawit Polkong
}

\address{Department of Physics, Faculty of Science, Mahidol University, 272 Rama VI Rd., Ratchathewi, Bangkok, 10400, Thailand}

\author{
Pitayuth Wongjun
}

\address{The Institute for Fundamental Study (IF), Naresuan University, 99 Moo 9, Tah Poe, Mueang Phitsanulok, Phitsanulok, 65000, Thailand}

\ead{suppawitpolkong@gmail.com}

\date{\today}

\begin{abstract}

The greybody factor of a massive scalar field in the Reissner-Nordstr\"om black hole is investigated using the Wentzel-Kramers-Brillouin (WKB) approximation and rigorous bound methods. 
We found that the transmission probability and behavior of the potential are directly related in such a way that the higher the potential, the lower the greybody factor. Both methods achieve a similar conclusion, which states that the graybody factor and the mass of the scalar field have an inverse relationship. This can be interpreted in a similar way in quantum mechanics, namely the scalar field with the higher mass will encounter a stronger interaction from the potential and then it is more difficult to penetrate through the potential barrier.
The rigorous bound has the advantage of not only being possible to calculate analytically, but also being applicable to a wider range of parameter values compared to the standard WKB approximation.

\end{abstract}

%\maketitle

%%%%%%%%%%%%%%%
\section{Introduction}
%%%%%%%%%%%%%%%

General Relativity (GR) provides the possibility of the existence of a mysterious object known as a black hole. With observational data \cite{Akiyama:2019cqa,TheLIGOScientific:2016src}, it suggests that black holes can exist in the real world. Moreover, the observational data provide a way to constrain black hole parameters characterized by any deviation from GR \cite{Garofalo:2020ajg,Vincent:2020dij,Dokuchaev:2020rye,Stepanian:2021vvk,Cornish:2017jml}. These provide the reason why the study of black holes receives much attention nowadays.

One of the most important characteristic behaviors is that black holes behave as thermal systems. Particularly, black holes carry entropy and can emit a type of radiation called Hawking radiation \cite{Hawking:1974sw,Hawking:1976de}. As a result, at the event horizon, the spectrum of the radiations from black holes is the same as that of the black-body spectrum. By taking the curvature of spacetime into account, the Hawking radiation is modified while propagating to spatial infinity. The greybody factor is defined as the ratio of the radiation at spatial infinity to that from the emitter by black holes. In fact, in terms of quantum mechanics, the greybody factor corresponds to the transmission amplitude of the wave where the curvature of the spacetime due to the black hole acts as the potential barrier.

The greybody factors from various kinds of spacetime geometry have been intensively investigated by various methods. One of the proper choices intensively investigated in literature is that of using WKB approximation \cite{Iyer:1986np,Parikh:1999mf,Cho:2004wj,Konoplya:2010kv,Dey:2018cws,Konoplya:2019ppy,Konoplya:2019hlu,Devi:2020uac}. It provides a good approximation for a simple form of spacetime geometries, and requires the high potential in order to obtain the enough precision in numerical calculation. This infers that the WKB method cannot be properly used for low potential which corresponds to a low multipole.
The other useful choice that allows us to study the behavior of the greybody factor analytically is to use the rigorous bound of the greybody factor instead of using the exact value of the greybody factor.
\cite{Visser:1998ke,Boonserm:2008qf,Boonserm:2009zba,Boonserm:2017qcq,Boonserm:2019mon,Barman:2019vst,Chowdhury:2020bdi,Kanzi:2020cyv}.

The studies on greybody factor are mainly conducted for massless scalar fields since it is instructive and supposed to provide stronger spectrum of the greybody factor. However, given mass to the scalar field is expected to provide a significant different behavior from the massless one. For example, it is found that there exist a critical mass which yields the maximum spectrum of the greybody factor \cite{Boonserm:2023oyt}. Moreover, the
mass of the scalar field can influence the instability of the scalar field itself \cite{Furuhashi:2004jk,Vieira:2021nha} and the superradiant instability \cite{Bekenstein:1998nt}. Furthermore, the massive scalar field plays a crucial role in fundamental theoretical physics as it can found from in the Kaluza-Klein models. Specifically, the behavior of a massless scalar field in the Fourier modes is similar to the massive one. Therefore, it is worthwhile to study the greybody factor contributed from the massive fields in order to gain insight into the behaviour of the thermal properties of the black hole through the effect of the mass of the field. 

In this paper, we are interested in investigating the greybody factor as the massive scalar field emitted from the charged black hole by using mainly WKB approximation and rigorous bound methods. Even though the investigation of a charged black hole is usually viewed as a simple toy model of the black hole, it provides an instructive study to a more realistic black hole. For example, the structure of the horizons is equivalent to the rotating black hole. Based on the charged black hole, the exact regular black hole solution can be obtained by considering the nonlinear electrodynamics coupled to Einstein gravity \cite{Ayon-Beato:1998hmi,Ayon-Beato:1999kuh,Ayon-Beato:2000mjt}. Moreover, the charged black hole including magnetic charge can provide an ability to catch fundamental theory associating with the existence of the magnetic monopole via the black hole perturbations \cite{Pereniguez:2023wxf,DeFelice:2023rra}.

By comparing the results from both methods, we found that
the bounds associated with our chosen functions are proper lower bounds of the greybody factor. The criteria of using the functions mentioned above to provide a suitable bound is discussed.
It is important to note that, even though we have not found the critical mass of a scalar field as found in literature, the scalar field with the higher mass will encounter a stronger interaction from the potential and then it is more difficult to penetrate through the potential barrier which is similar to one in quantum mechanics.
We also found that the advantage of the rigorous bound is that it is not only an analytical form, but also can be used
for a wider range of parameter values (compared to the WKB approximation method).

This paper is organized as follows.
In Sec.~\ref{sec:RW}, the Schr\"odinger-like equation for the radial part of the considered massive scalar field is derived.
The various methods of solving the aforementioned equation for the greybody factor are discussed and the results are reported in Sec.~\ref{sec:wkb} for the WKB approximation and Sec.~\ref{sec:bound} for the rigorous bound methods.
Sec.~\ref{sec:conclu} is devoted to conclusions.

%\newpage
%%%%%%%%%%%%%%%%%%%%%%%%%%%%%
\section{Regge-Wheeler equation}
\label{sec:RW}
%%%%%%%%%%%%%%%%%%%%%%%%%%%%%

Let us consider a massive scalar field $\Phi(x^\mu)$ on the static and spherically symmetric spacetime which is simply described by the form of the line element: 
\ba
	\dd s^2=-f(R)\dd t^2+\frac{1}{f(R)}\dd R^2+R^2\Big(\dd\theta^2+\sin^2\theta\dd\phi^2\Big),
\ea
where $f(R)$ is the horizon function. %$\dd\Omega^2=\dd\theta^2+\sin^2\theta\dd\phi^2$
Note that this form of metric is obtained from the specific condition on the components of the energy-momentum tensor $T^{\,t}_{\,\,\,\,\,t}=T^{\,R}_{\,\,\,\,\,R}.$
The dynamics of the field obey the Klein-Gordon equation:
\ba
	{\frac {1}{\sqrt {-g}}}\partial _{\mu }\left(g^{\mu \nu }{\sqrt {-g}}\pa_\nu\Phi\right)-\Mph^2\Phi=0,\label{KG eq}
\ea
where $\Mph$ is the mass of the field.
Employing the separation of variable method and imposing the form of the field given by
\ba
	\Phi(t, R, \theta,\phi)=T(t)\frac{\psi(R)}{R}\Theta(\theta,\phi).
\ea
Plugging it into Eq.~\eqref{KG eq}, one finds that the forms of the temporal and angular parts satisfying their own equations are, respectively, expressed as
\ba
	T(t)=e^{i\Om t},
	\qquad
	\Theta(\theta,\phi)=Y_{lm}(\theta,\phi),
\ea
where $\Om$ is a constant and $Y_{lm}(\theta,\phi)$ is the spherical harmonics with a non-negative integer $l$ and integer $m$ where $l\geq|m|$.
As a result, the leftover is the radial part which can be written in the Schr\"odinger-like equation as 
\ba
	\frac{\dd^2 \psi(R)}{\dd R^{*2}}+\lt[\Om^2-V(R)\rt]\psi(R)=0,\label{RW eq}
\ea
with the potential $V(R)$ given by
\ba
	V(R)=f(R)\lt[\frac{l(l+1)}{R^2}+\frac{f'(R)}{R}+\Mph^2\rt].\label{eff pot}
\ea
Here, the tortoise coordinate $R^*$ is introduced as $R^*=\int |f(R)|^{-1}\dd R$. 
In this coordinate, the black hole's outer horizon $R_h$ is mapped to negative infinity $R^*\to-\infty$.
The prime denotes the derivative with respect to the radial coordinate.
Eq.~\eqref{RW eq} also known as the Regge-Wheeler equation describes the dynamics of the field $\psi$ through the potential barrier due to spacetime curvature with frequency (or energy) $\om$.

In this study, we are interested in the charged black hole described by the horizon function:
\ba
	f(R)=1-\frac{2GM}{R}+\frac{GQ^2}{R^2},\label{hor func}
\ea
where $M, Q$ and $G$ are the mass of the black hole, electric charge, and Newton's gravitational constant, respectively.
Note that the outer horizon can be determined as the largest root of $f(R_h)=0$.
It is obtained as $R_h=GM+\sqrt{G^2M^2-GQ^2}$.
For convenience, let us work with the dimensionless variables by rescaling the variables with $R_h$ as follows:
\ba
	r=\frac{R}{R_h},\qquad
	q=\frac{\sqrt{G}Q}{R_h},\qquad 
    m=\frac{GM}{R_h}.
\ea
The dimensionless variables in the perturbation equation can be defined by
\ba
    \om=R_h\Om,\qquad
	\mph=R_h\Mph,\qquad
	v(r)=R_h^2V(R).
\ea
As a result, the dimensionless versions of the Regge-Wheeler equation~\eqref{RW eq} and potential in Eq.~\eqref{eff pot} are, respectively, expressed as
\ba
	\frac{\dd^2 \psi(r)}{\dd r^{*2}}+\lt[\om^2-v(r)\rt]\psi(r)=0,\label{RW eq2}
\ea
and 
\ba
    v(r)&=&\lt(1-\frac{2m}{r}+\frac{q^2}{r^2}\rt)\lt[\frac{l(l+1)}{r^2}+\frac{2m}{r^3}-\frac{2q^2}{r^4}+\mph^2\rt].\label{v}
\ea
From the fact that the outer horizon radius depends on both black hole's mass and charge, we are interested in analyzing how the existence of the electric charge affects the behavior of the potential.
Therefore, in our consideration, the mass can change in such a way that the black hole's horizon radius is fixed, i.e.,
\ba
    m&=&\frac{1+q^2}{2}.\label{m in q}
\ea
Note that the valid value of the charge lies in the range $0\leq q\leq1$ where $q=1$ corresponds to the extremal limit (the inner and outer horizons coincide).
One can check that $f(r)$ and $f'(r)$ are always positive outside the black hole ($r>1$).
Consequently, the potential $v(r)$ outside the black hole horizon is positive.
In addition, the asymptotic values of the potential at the black hole's outer horizon and spatial infinity can be, respectively, determined as
\ba
    v(r^*\to-\infty)
    \,\,\,=&v\lt(r\to1\rt)
    &=\,\,\,0,\\
    v(r^*\to\infty)
    \hspace{0.5cm}=&v(r\to\infty)
    &=\,\,\,\mph^2.
\ea
These values are indeed implied from the vanishing and flatness behaviors at the horizon and asymptotically far distance of the horizon function in Eq.~\eqref{hor func}, respectively.

\begin{figure}[h]
\begin{center}
\includegraphics[scale=0.52]{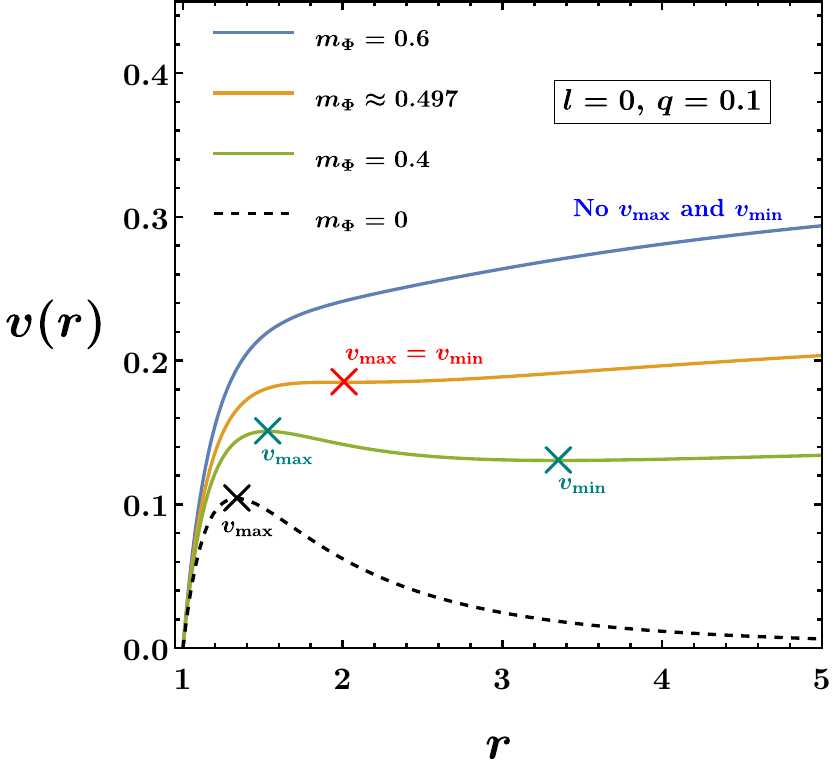}
\hspace{1cm}
\includegraphics[scale=0.52]{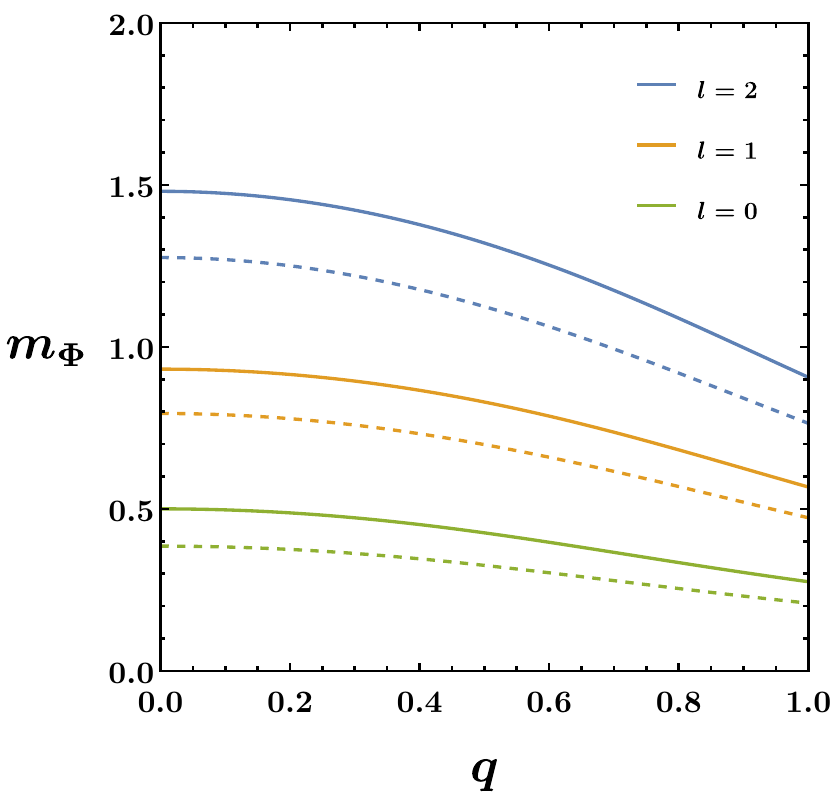}
\end{center}
\vspace{-0.5cm}
\caption{Left: the potential for various values of $\mph$ corresponding to the cases that local extrema exist and disappear.
Right: the solid curves represent the values of $\mph$ at which the local extrema of the potential merge as a single point. The dashed curves represent the values of $\mph$ that satisfy $\vm=\mph^2$.
}\label{fig:v vary mph}
\end{figure}

Let us first consider how the change in $\mph$ affects the behavior of the potential.
When the perturbed field is massless ($m_\Phi=0$), the potential behaves as a barrier-like one with a single maximum point (denoted as $\vm$).
On the other hand, for the massive scenario ($m_\Phi\neq0$), the potential does not vanish at asymptotic infinity.
A local minimum point of the potential (denoted as $\vmin$) can emerge when $\mph$ is sufficiently small.
This minimum point is located at a distance greater than the distance corresponding to $\vm$, $r_{\vmin}>r_{\vm}$.
As $\mph$ increases, the maximum and minimum points get closer and coincide at a certain value of $\mph$.
When $\mph$ exceeds the aforementioned value, there are local extremum points.
In other words, the potential in this regime is a monotonically increasing function in $r$ that grows from zero to $\mph^2$.
It is also found that the value of $v(r)$ at its local minimum point is always less that its value at asymptotic infinity, i.e.,  $\vmin<\mph^2$.
However, $\vm$ can be greater than, less than or equal to $\mph^2$.
For example of the case $l=0$ and $q=0.1$ (see the left panel of Fig.~\ref{fig:v vary mph}), the potential satisfies the conditions $\vm=\mph^2$ and $\vm=\vmin$ when $\mph\approx0.382$ and $\mph\approx0.497$, respectively.
Furthermore, the potential is higher as $\mph$ increases.
There are certain features of $v(r)$ that are true for all possible values of $l$ and $q$.
The right panel of Fig.~\ref{fig:v vary mph} shows the value of $\mph$ associated with the conditions $\vm=\mph^2$ (dashed curves) and $\vm=\vmin$ (solid curves).
This means that the potential with $\mph$ below the dashed curve will have both the local minimum point and the maximum point (with $\vm>\mph^2$).
For the region of $\mph$ between the dashed and solid curves, there are also both local extremum points, but $\vm<\mph^2$.
For the region of $\mph$ above the solid curve, there are no local extremum points.
Another remark is that the potential behaves similarly even in the Schwarzschild black hole case ($q=0$) and the extremal black hole case ($q=1$). 

\begin{figure}[h]
\begin{center}
\includegraphics[scale=0.52]{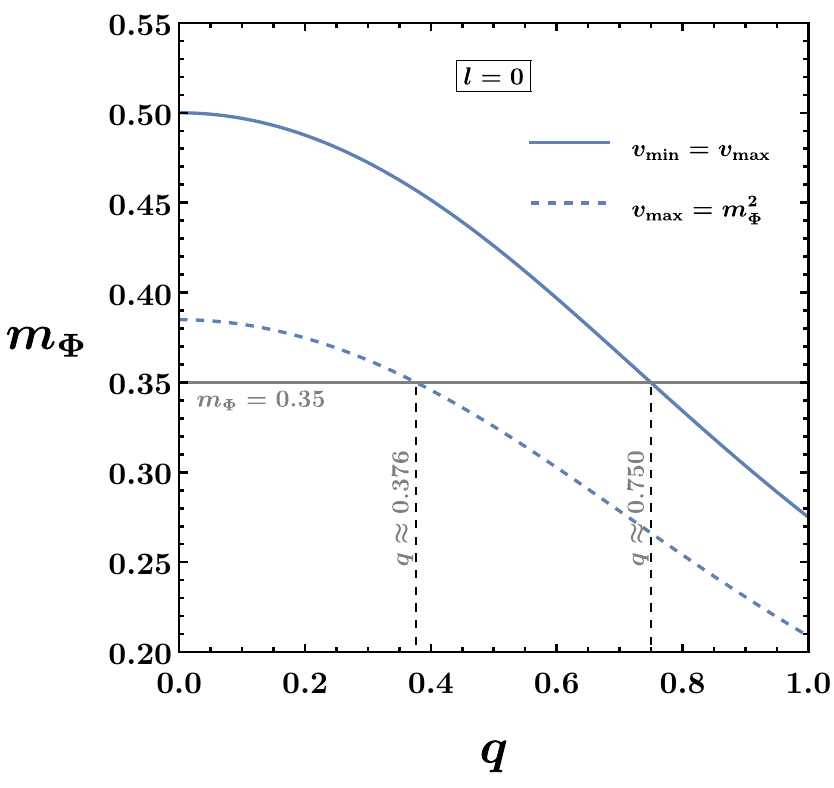}
\hspace{1cm}
\includegraphics[scale=0.525]{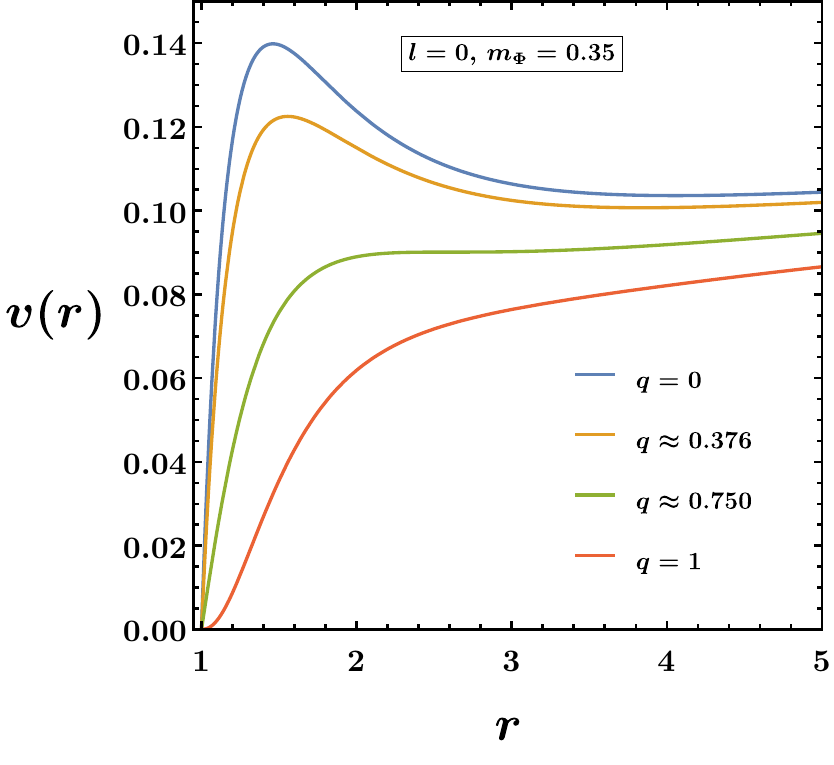}
\end{center}
\vspace{-0.5cm}
\caption{
Left: the values of $\mph$ at which $\vm=\vmin$ (solid curve) and $\vm=\mph^2$ (dashed curve).
Right: the potential for various values of $q$ with fixing $l=0$ and $\mph=0.35$.
}\label{fig:v vary q}
\end{figure}

\begin{figure}[h]
\begin{center}
\includegraphics[scale=0.35]{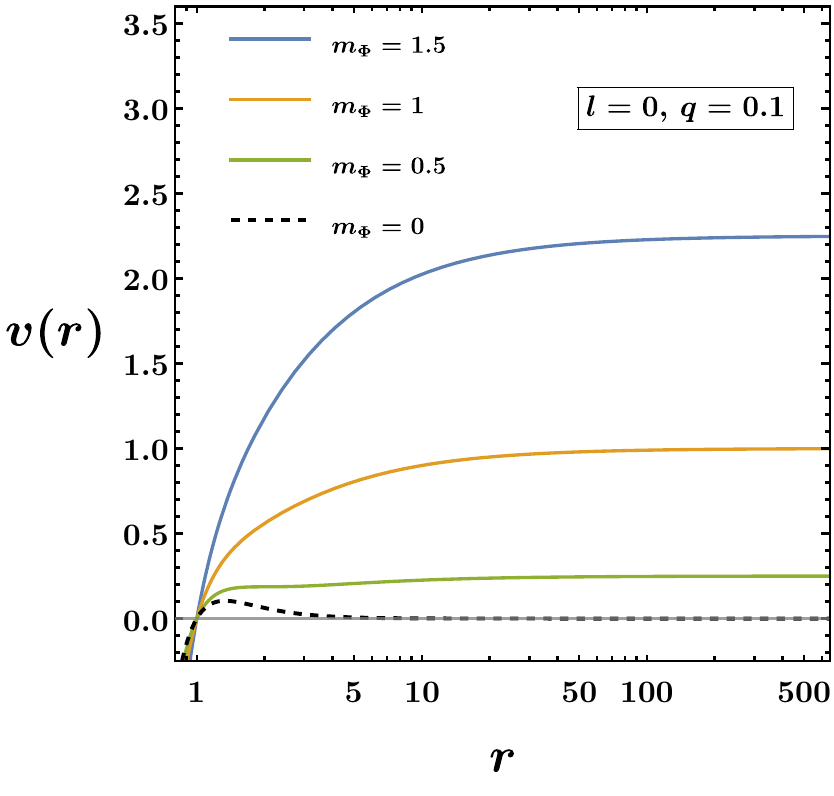}
\hspace{0.3cm}
\includegraphics[scale=0.35]{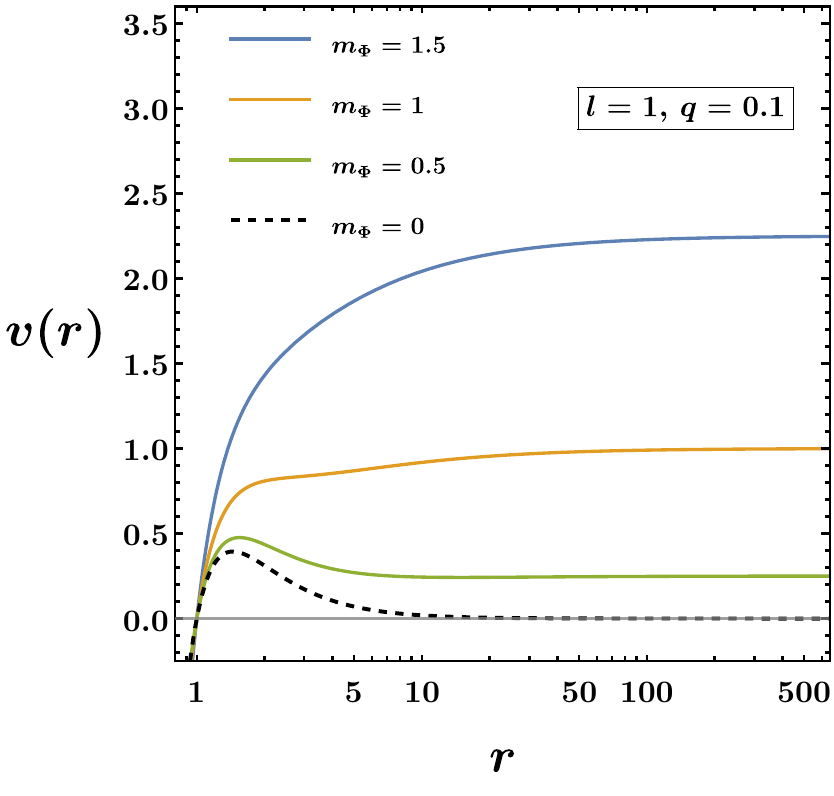}
\hspace{0.3cm}
\includegraphics[scale=0.35]{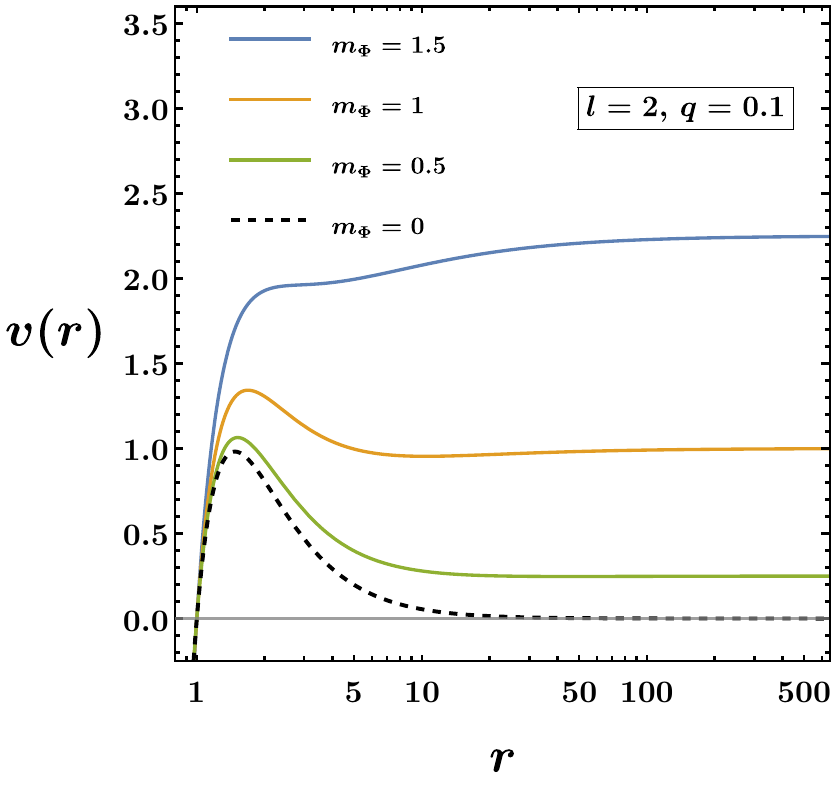}
\\\vspace{0.3cm}
\includegraphics[scale=0.35]{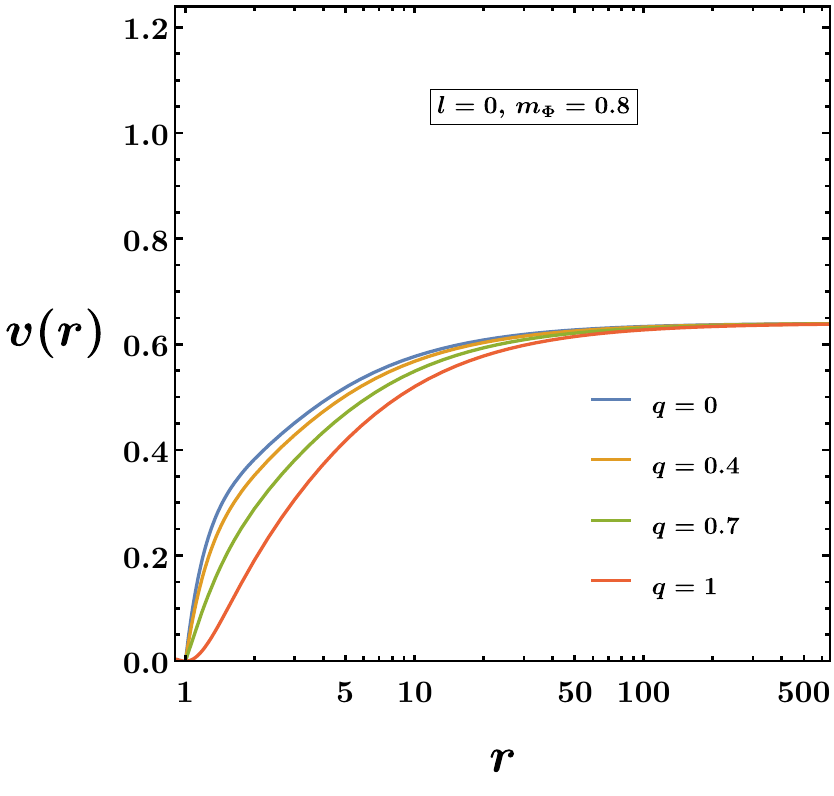}
\hspace{0.3cm}
\includegraphics[scale=0.35]{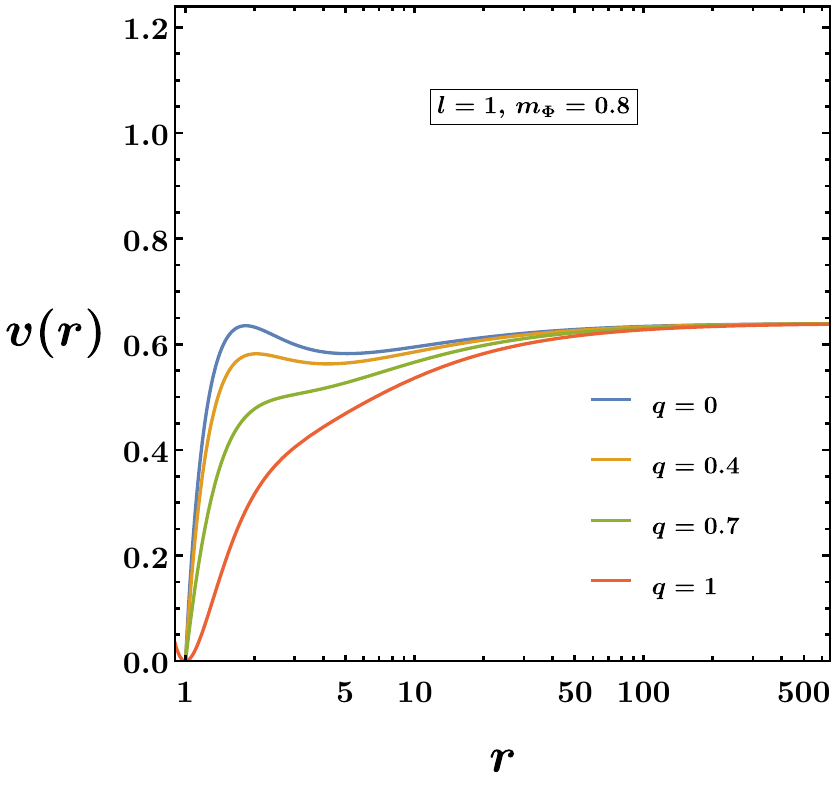}
\hspace{0.3cm}
\includegraphics[scale=0.35]{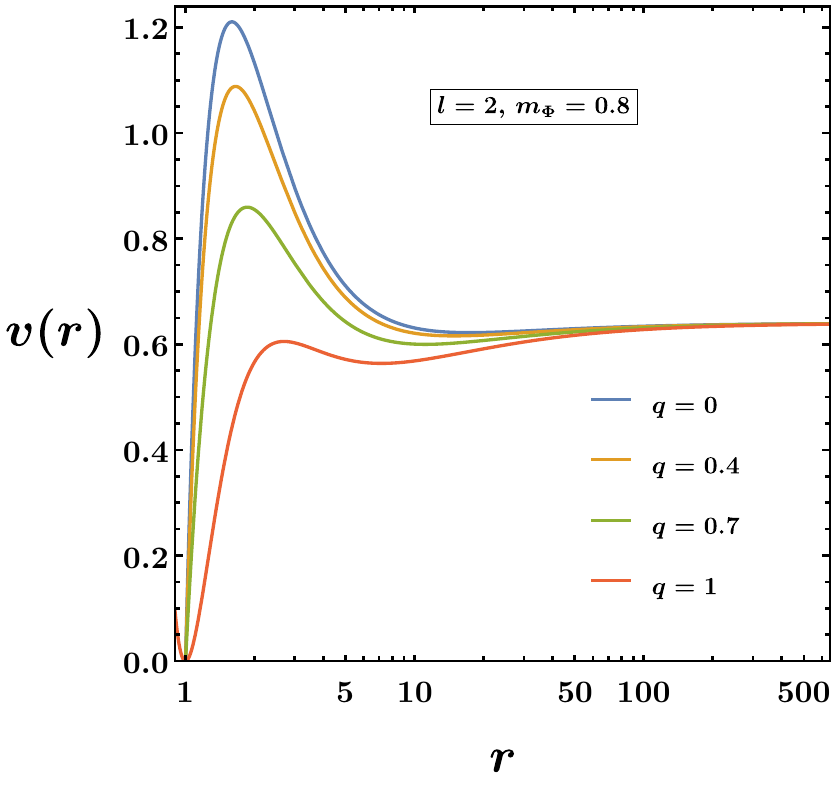}
\end{center}
\vspace{-0.5cm}
\caption{
The behavior of the potential for varying $\mph$ (top panels) and varying $q$ (bottom panels). The first, second and third columns show the potentials corresponding to $l=0$, $l=1$ and $l=2$, respectively.
}\label{fig:v vary l}
\end{figure}

The effect of $q$ on the potential is now discussed.
Unlike the effect of $\mph$, the behavior of the potential does not change significantly with varying $q$ if $\mph$ is too small or large.
For example, the potential with $l=0$ always has local extremum points and does not have those points when $\mph>0.5$ and $\mph\lessapprox0.275$, respectively.
When $\mph$ lies in the intermediate range: $0.275\lessapprox\mph<0.5$, the emergence of local extremal points can be affected by the variation of $q$.
These features can be illustrated in Fig.~\ref{fig:v vary q} for the case of $l=0$ and $\mph=0.35$.
There exist local extrema points and there do not exist those points when $q\lessapprox0.750$ and $q\gtrapprox0.750$, respectively.
In summary, the contribution due to $q$ affects $v(r)$ in such a way that the larger the charge, the lower the potential.
Another parameter that affects the potential is $l$.
As can be clearly seen in Eq.~\eqref{v}, the potential is higher as $l$ increases.
$l$ provides a stronger contribution to the potential at a shorter distance.
It can be illustrated in Fig.~\ref{fig:v vary l}.

So far, one can see that the curvature of spacetime plays a role as a potential barrier.
The radial part of the wave function corresponding to the perturbed scalar field penetrates through such a potential similar to the particle in the potential well in quantum mechanics.
In the next sections, we aim to determine the greybody factor, which is equivalent to the transmission probability using the WKB approximation and rigorous bound methods.

%%%%%%%%%%%%%%%%%%%%%%%%%%%%%%%%%%%%%
\section{Greybody factors from WKB approximation method}
\label{sec:wkb}
%%%%%%%%%%%%%%%%%%%%%%%%%%%%%%%%%%%%%

The key idea of the WKB approximation is the use of series expansion to obtain the solution. The coefficients of series expansion can be obtained by matching the solutions at the boundaries in which the frequency of the particle is equal to the potential. Therefore, in order to perform the calculation properly, one can separate the consideration into three cases; low-frequency, intermediate-frequency, and high-frequency regimes of the particle. For the high-frequency case, most of the particles can penetrate through the potential so that it is a trivial case. We will omit the consideration for this case in this article. It is convenient to redefine the Regge-Wheeler equation~\eqref{RW eq2} as follows:
\begin{equation}
\left(\frac{\dd^{2}}{\dd {r^{*}}^{2}}+Q\right)\psi=0,
\end{equation}
where $Q=\omega^{2}-v(r)$.

\subsection{Intermediate-frequency approximation ($\omega^{2}\approx \vm$)}
For the intermediate frequency approximation, one cannot apply the WKB method directly since the solutions cannot be matched at the boundary. In order to overcome such obstruction, one can expand the potential around the maximum point for the solution evaluated in the region $\omega^{2} < \vm$ \cite{Iyer:1986np,Parikh:1999mf,Cho:2004wj,Konoplya:2010kv}. By using this idea and following the procedure found in  \cite{Cho:2004wj}, the approximate greybody factor for intermediate-frequency regime ($T_\text{IWKB}$) can be obtained as follows:
\begin{equation}
T_\text{IWKB} = \frac{1}{1+\exp\big[2S(\om)\big]},
\end{equation}
where the function $S$ is written as 
\begin{eqnarray}
    S(\omega) 
    =\hspace{-0.6cm}
    &&\pi k^{\frac{1}{2}}\lt[
    \frac{1}{2}z_{0}^{2}
    +\lt(\frac{15}{64}b^{2}_{3}-\frac{3}{16}b_{4}\rt)z^{4}_{0}
    +\lt(\frac{1155}{2048}b^{4}_{3}-\frac{315}{256}b^{2}_{3}b_{4}+\frac{35}{128}b_{4}^{2}
    +\frac{35}{64}b_{3}b_{5}-\frac{5}{32}b_{6}\rt)z^{6}_{0}
    \rt]
    \nonumber\\ 
    &&+\frac{\pi}{ k^{\frac{1}{2}}}
    \lt[\lt(\frac{3}{16}b_{4}-\frac{7}{64}b_{3}^{2}\rt)
    -\lt(\frac{1365}{2048}b^{4}_{3}-\frac{525}{256}b^{2}_{3}b_{4}
+\frac{85}{128}b^{2}_{4}+\frac{95}{64}b_{3}b_{5}-\frac{25}{32}b_{6}\rt)z^{2}_{0}\rt]+\mathcal{O}(\omega).\hspace{0.8cm}
\end{eqnarray}
Note that short-hand variables evaluated at the maximum of the potential can be defined as follows:
\begin{equation}
z^{2}_{0}= -\frac{Q_{max}}{k}, \quad 
k = \frac{1}{2}\frac{\dd^{2}Q}{\dd x^{2}},\quad 
b_{n}=  \frac{1}{n!k} \frac{\dd^{n}Q }{\dd x^{n}},
\end{equation}
where $\mathcal{O}(\omega)$ denotes the higher-order terms.

\subsection{Low-frequency approximation ($\omega^{2}\ll  \vm$)}

In this case, the classical turning points where $\omega^2 = v$ will be much farther apart compared to the previous case and denoted by $r_1$ (smaller radius) and $r_2$ (larger radius). By using the low-frequency approximation, the greybody factor $T_\text{LWKB}$ can be expressed as \cite{Cho:2004wj}
\begin{equation}
T_\text{LWKB} = \exp\lt(-2\int_{r_1}^{r_2} \sqrt{v-\omega^{2}}\,\dd r^*\rt)
= \exp\lt(-2\int_{r_1}^{r_2} \frac{\sqrt{v-\omega^{2}}}{f}\,\dd r\rt).
\end{equation}
From this expression, one can see that the turning points can be obtained by specifying the value of $\omega$. Therefore, we need to evaluate the greybody factor point by point. Moreover, since the potential is complicated and the integration cannot be analytically performed, we will use the numerical method for this case.

\subsection{Results}

\begin{figure}[h]
\begin{center}
\includegraphics[scale=0.5]{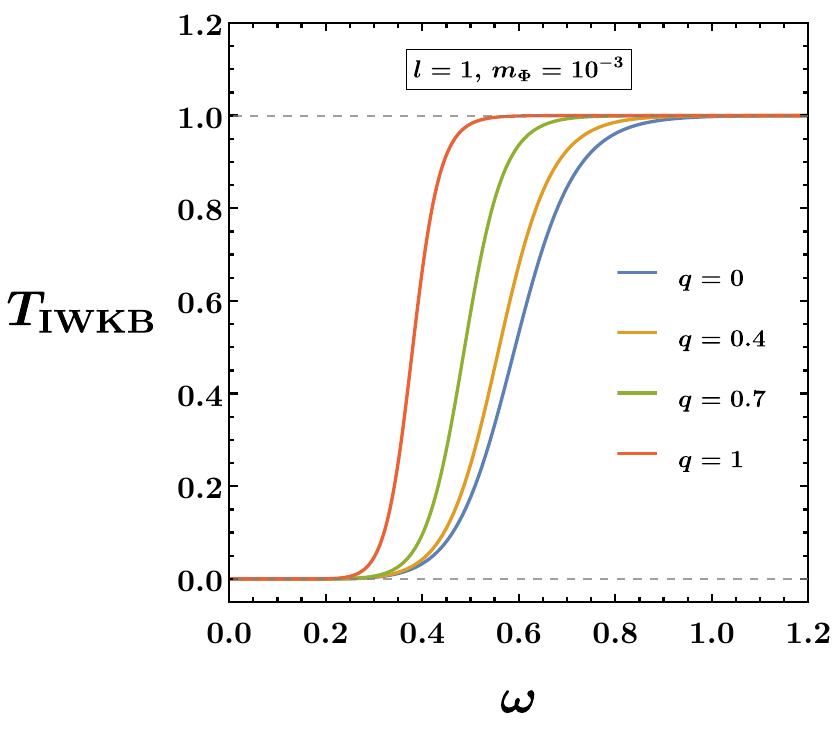}
\hspace{1cm}
\includegraphics[scale=0.5]{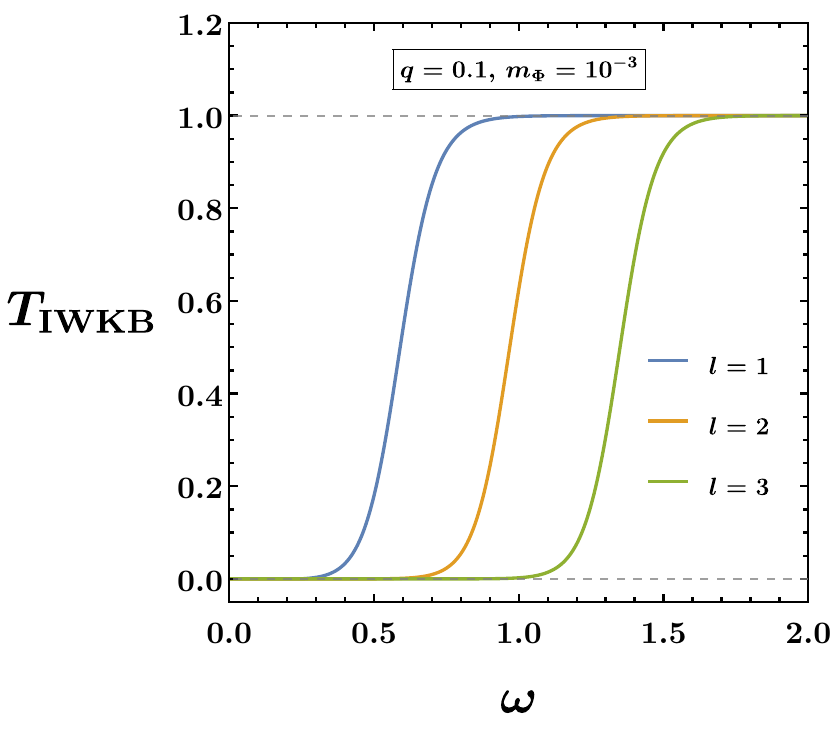}
\\
\hspace{.5cm}
\includegraphics[scale=0.5]{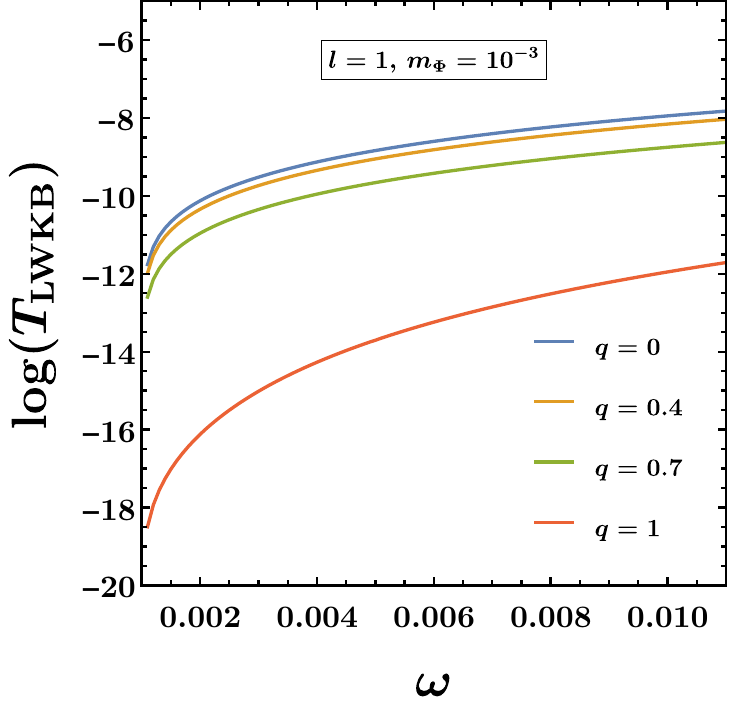}
\hspace{1.9cm}
\includegraphics[scale=0.5]{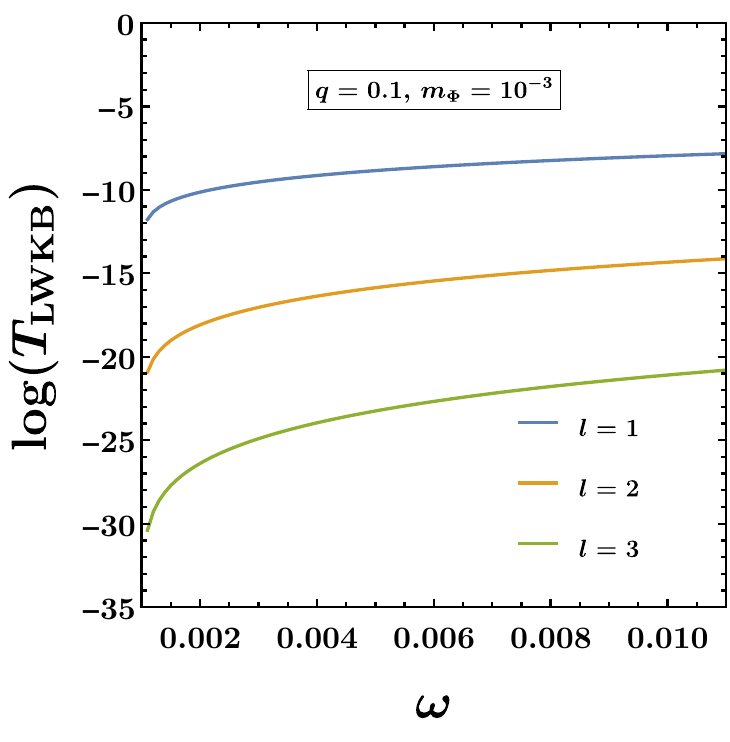}
\end{center}
\vspace{-0.5cm}
\caption{Greybody factor from WKB approximation for both intermediate frequency (IWKB, top panels) and low frequency (LWKB, bottom panels) regimes. 
}\label{fig:Twkb}
\end{figure}

According to the assumption of the WKB approximation method, the potential is required to behave as a barrier-like one.
Hence, the potential at $\vm$ must be sufficiently larger than its value at asymptotic infinity, i.e., $\mph^2$.
To guarantee that the WKB is applicable, $\mph$ is chosen as $10^{-3}$ and we consider $\om\geq1.1\times10^{-3}$. 
The results of the greybody factors for various scenarios can be illustrated in Fig.~\ref{fig:Twkb}.

By comparing the greybody factor to the shape of the potential, it is found that the results from the WKB approximation are in the same trend as found in quantum mechanics. Specifically, the higher the potential, the lower the greybody factor. %This can be illustrated in Fig.~\ref{fig:TWKBl1}. 
Note that the results of the WKB approximation provide a good estimate for the high value of $l$ and the lower value of $m_\Phi$. This is due to the fact that at the lower value of $l$ or higher value of $m_\Phi$, the turning points of the potential are far from together. Therefore, expansion of the potential around the maximum may not be sufficiently valid. This is the reason why we choose $l\geq1$ for the plots in Fig.~\ref{fig:Twkb} even though the greybody factors of these modes are much lower than one for the $l=0$ mode.

%%%%%%%%%%%%%%%%%%%%%%%%%%%%%%%%%%%%%
\section{Greybody factors from rigorous bound method}
\label{sec:bound}
%%%%%%%%%%%%%%%%%%%%%%%%%%%%%%%%%%%%%

One of the other treatments in determining the greybody factor is the so-called rigorous bound method \cite{Boonserm:2008qf,Boonserm:2009zba}.
This method provides the lower bound of the greybody factor.
The advantage is the bound can be obtained in analytic form, unlike the results from the WKB method discussed in the previous section.
Based on Refs.~\cite{Boonserm:2008qf,Boonserm:2009zba}, the lower bound of the greybody factor is given by 
\ba
	T\geq\text{sech}^2\lt(\int_{-\infty}^\infty\vartheta\,\dd r^*\rt),\label{bound gen}
\ea
where
\ba
	\vartheta=\frac{1}{2h(r^*)}\sqrt{h'(r^*)^2+\Big[\om^2-v(r)-h(r^*)^2\Big]^2},
\ea
for some positive function $h(r^*)$ satisfying
\ba
	h(\pm\infty)=\sqrt{\omega^2-v(\pm\infty)}.
\ea 
In this study, we proposed two possible forms of the function $h(r^*)$ in investigating the bound of the greybody factor.

\subsection{$h=\sqrt{\omega^2-v}$}
The bound of the greybody factor in Eq.~\eqref{bound gen} can be written as
\ba
	T
	\geq
	T_\text{bound1}
	=\text{sech}^2\lt(\frac{1}{2}\int_{-\infty}^\infty\frac{|h'|}{h}\dd r^*\rt),\label{Tb h1}
\ea
where the prime denotes the derivative with respect to the tortoise coordinate $r^*$.
Let us start with the large-mass regime so that the potential is monotonically growing in $r^*$.
It implies that $v'>0$.
One then knows that the derivative of the function $h$ for this regime is always negative since
\ba
	h'=-\frac{v'}{2h}.
\ea
The bound of the greybody factor  Eq.~\eqref{Tb h1} can be computed as
\ba
	T_\text{bound1, (large $\mph$)}
%&=&
=
\text{sech}^2\lt(-\int_{-\infty}^{\infty}\frac{h'}{2h}\dd r^*\rt)
%\no\\		
%&=&\text{sech}^2\lt[\frac{1}{2}\lt(-\ln(h)\Big|_{-\infty}^{\infty}\rt)\rt]\no\\
%&=&\text{sech}^2\lt[-\ln\lt(\sqrt{\frac{h_\infty}{h_{-\infty}}}\rt)\rt]\no\\
%&=&\frac{4h_{-\infty}h_\infty}{\lt(h_{-\infty}+h_\infty\rt)^2}\no\\
%&=&
=
\frac{4\om\sqrt{\om^2-\mph^2}}{\lt(\om+\sqrt{\om^2-\mph^2}\,\rt)^2}.
\label{Tb1 large m}
\ea
The property: $\text{sech}\big[\ln(x)\big]=2x/(1+x^2)$ is also used in the above calculation.
According to the above result, this bound exists for the frequency satisfies $\om\geq\mph$.
It is surprising that the bound in such a regime depends only on the perturbed field's mass.

For the small-mass regime, the behaviors of the function $h$ can be split into three ranges of $r^*$ as follows: 
$-\infty<r^*<r_{\vm}$, 
$r_{\vm}<r^*<r_{\vmin}$,
and $r_{\vmin}<r^*<\infty$.
The bound in Eq.~\eqref{Tb h1} for this regime can be computed as
\ba
	T_\text{bound1, (small $\mph$)}
&=&\text{sech}^2\lt(-\frac{1}{2}\int_{-\infty}^{\vm}\frac{h'}{h}\dd r^*+\frac{1}{2}\int_{\vm}^{\vmin}\frac{h'}{h}\dd r^*-\frac{1}{2}\int_{\vmin}^\infty\frac{h'}{h}\dd r^*\rt)\no\\			
%&=&\text{sech}^2\lt[\frac{1}{2}\lt(-\ln(h)\Big|_{-\infty}^{r^*_{\vm}}+\ln(h)\Big|_{r^*_{\vm}}^{r^*_{\vmin}}-\ln(h)\Big|_{r^*_{\vmin}}^\infty\rt)\rt]\no\\
%&=&\text{sech}^2\lt[-\ln\lt(\frac{h_{\vm}}{h_{\vmin}}\sqrt{\frac{h_\infty}{h_{-\infty}}}\,\rt)\rt]\no\\
%&=&\frac{4h_{\vm}^2h_{\vmin}^2h_{-\infty}h_\infty}{\lt(h_{\vm}^2h_{-\infty}+h_{\vmin}^2h_\infty\rt)^2}\no\\	
&=&\frac{4\om\sqrt{\om^2-\mph^2}\lt(\omega^2-\vm\rt)\lt(\omega^2-\vmin\rt)}{\lt[\lt(\om^2-\vm\rt)\sqrt{\om^2-\mph^2}+\om\lt(\om^2-\vmin\rt)\rt]^2}.
\label{Tb1 small m}
\ea
It is easily checked that the bound in Eq.~\eqref{Tb1 large m} can be recovered by taking $\vm=\vmin$ in Eq.~\eqref{Tb1 small m}.
When $\vm>\mph^2$, the above expression seems to be negative in the range $\sqrt{\vmin}<\om<\sqrt{\vm}$.
However, the condition $\om<\sqrt{\vm}$ makes the function $h=\sqrt{\om^2-v}$ not well-defined.
The bound of the greybody factor in Eq.~\eqref{Tb1 small m} is, therefore, valid only in the range $\om>\sqrt{\vm}\,(>\mph)$.
On the other hand, when $\vm<\mph^2$, the the bound exists in the same region of $\om$ for $T_\text{bound, (large $\mph$)}$, i.e., $\om\geq\mph$. 
It is worthwhile to note that the lower bound of the greybody factor $T_\text{bound}$ depends only on $\vm$ and $\vmin$, not explicitly on the shape of the potential.
Hence, it might be difficult to analyze the profile of the bound.

%%%%%%%%%%%%%%%%%%%%%%%%%%%%%%
\subsection{$h=\sqrt{\omega^2-f\mph^2}$}
\begin{figure}[h]
\begin{center}
\includegraphics[scale=0.35]{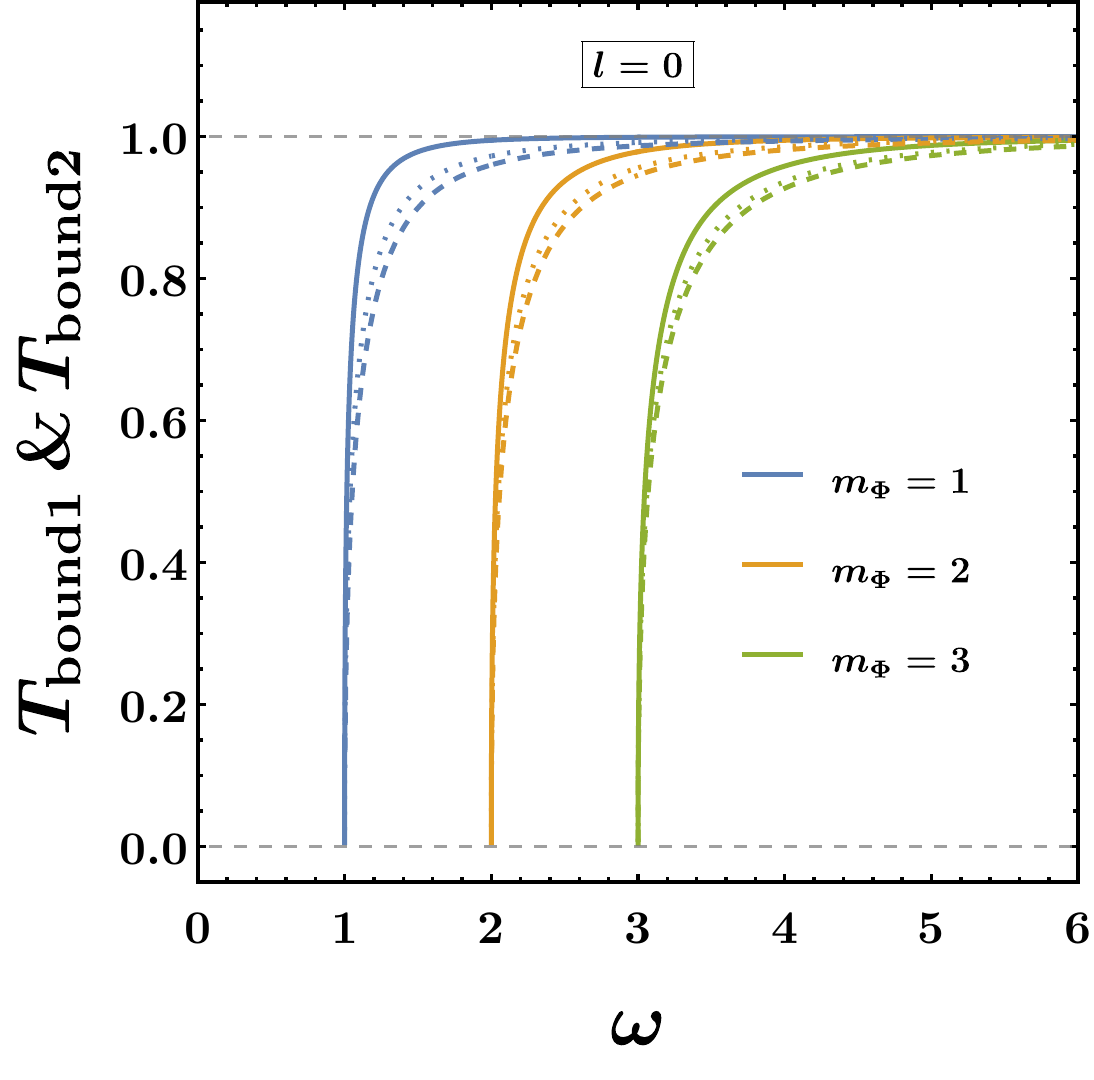}
\qquad
\includegraphics[scale=0.35]{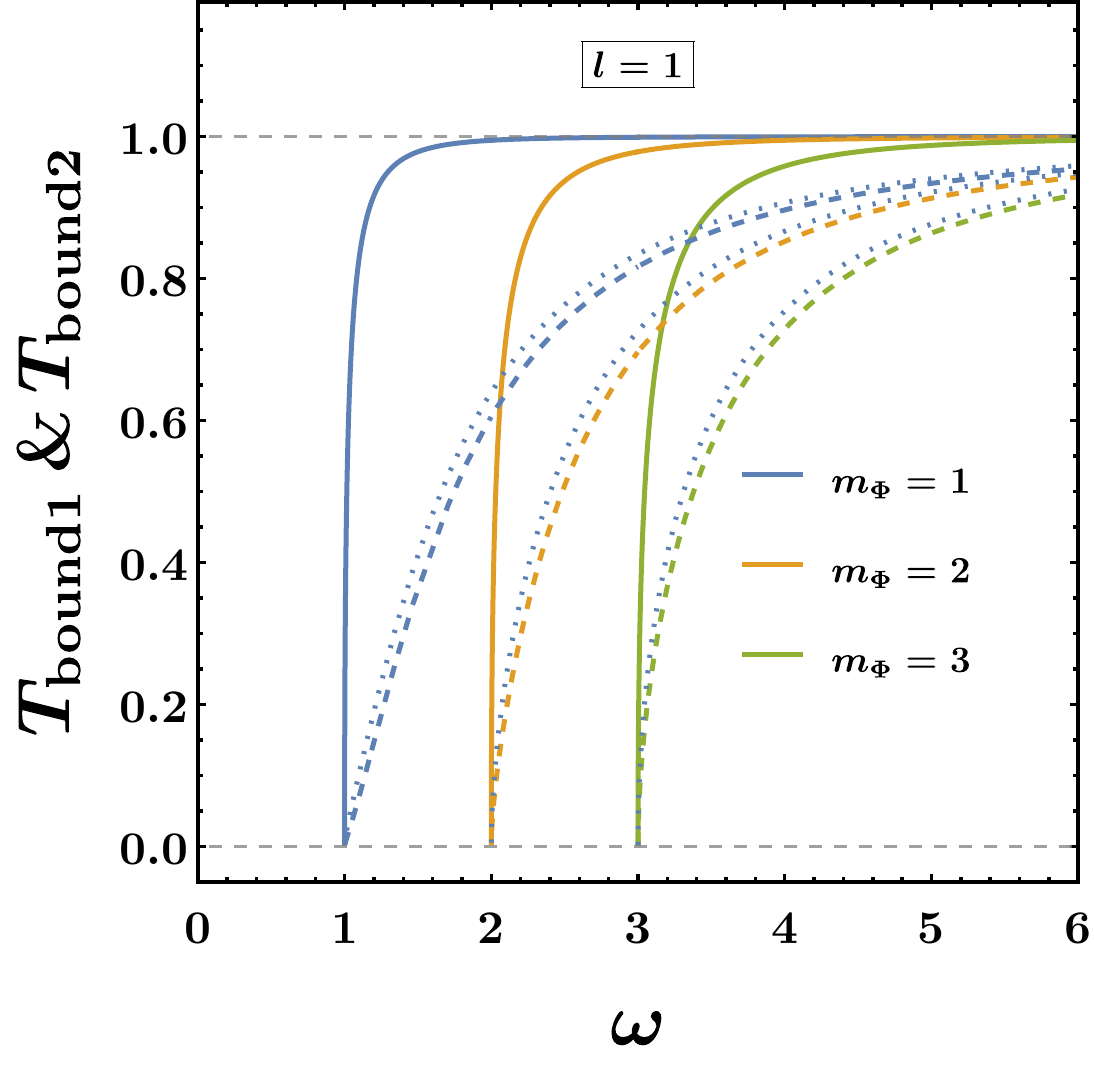}
\\
\includegraphics[scale=0.54]{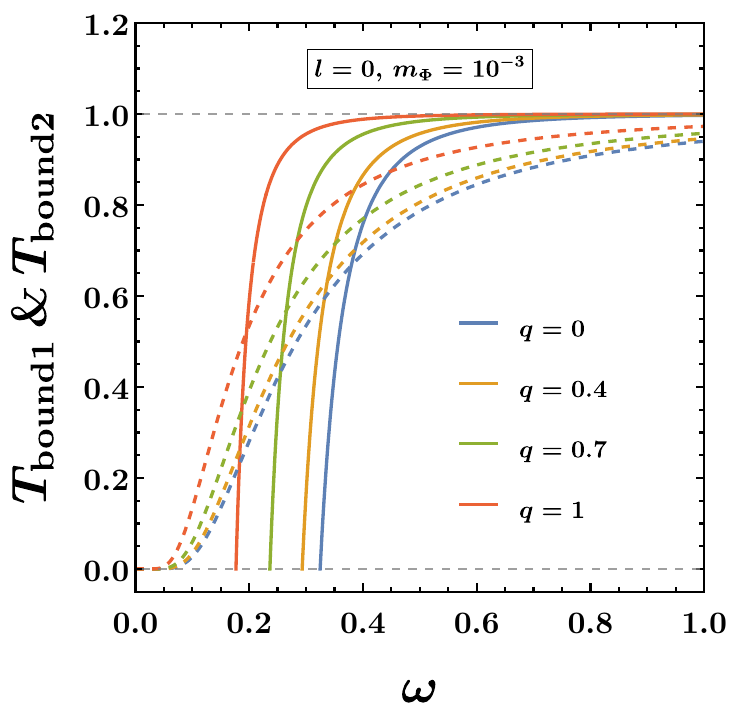}
\qquad
\includegraphics[scale=0.52]{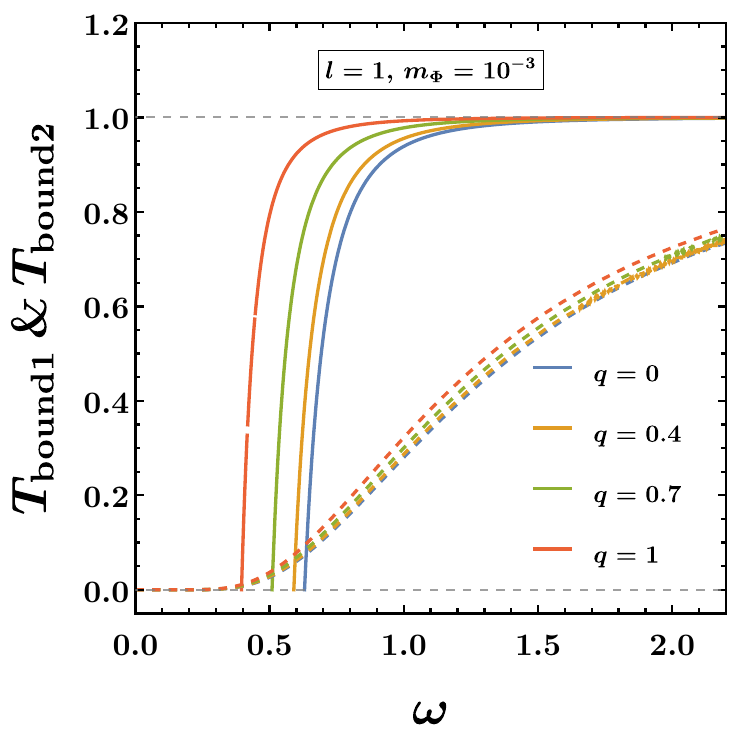}
\end{center}
\vspace{-0.5cm}
\caption{
Lower bounds of greybody factor associated with $h=\sqrt{\om^2-v}$ represented as solid lines and $h=\sqrt{\om^2-f\mph^2}$ represented as dotted and dashed lines. 
The top panels are the plots for large $\mph$ regime where the dotted and dashed lines correspond to $q=0.9$ and $q=0.1$, respectively.
The bottom panels are plots for small $\mph$ regime.
}\label{fig:Tb} 
\end{figure}

The bound of the greybody factor in Eq.~\eqref{bound gen} for this choice of the function $h$ can be expressed as follows:
\ba
	T
	&\geq&
    \text{sech}^2\lt(\frac{1}{2}\int_{-\infty}^\infty\frac{1}{h}\sqrt{h'^2+\big(\om^2-v-h^2\big)^2}\dd r^*\rt)\no\\
    &\geq&
    \text{sech}^2\lt(\frac{1}{2}\int_{-\infty}^\infty\frac{1}{h}\sqrt{\Big[h'+\big(\om^2-v-h^2\big)\Big]^2}\dd r^*\rt)\no\\
%&&\red{\text{It is okay because $h'\big(\om^2-v-h^2\big)>0$}}\no\\
%&=&\text{sech}^2\lt(\frac{1}{2}\int_{-\infty}^\infty\frac{1}{h}\Big|h'+\big(\om^2-v-h^2\big)\Big|\dd r^*\rt)\no\\
    &\geq&
    \text{sech}^2\lt(\frac{1}{2}\int_{-\infty}^\infty\frac{1}{h}\Big(|h'|+\big|\om^2-v-h^2\big|\Big)\dd r^*\rt)\no\\
%&&\red{\text{Use the triangle inequality $|x+y|\leq|x|+|y|$}}\no\\
%&=&\text{sech}^2\lt(\int_{-\infty}^\infty\frac{|h'|}{2h}\dd r^*+\int_{-\infty}^\infty\frac{\big|\om^2-v-h^2\big|}{2h}\dd r^*\rt)\no\\
%&=&\text{sech}^2\lt(\int_{-\infty}^\infty\frac{|h'|}{2h}\dd r^*+\int_{-\infty}^\infty\frac{\big|f\mph^2-v\big|}{2h}\dd r^*\rt)\no\\
%&=&\text{sech}^2\lt(-\int_{-\infty}^\infty\frac{h'}{2h}\dd r^*-\int_{-\infty}^\infty\frac{f\mph^2-v}{2h}\dd r^*\rt)\no\\
%&&\red{\text{$h'$ and $f\mph^2-v$ are always neg}}\no\\	
%&=&\text{sech}^2\lt(\int_{-\infty}^\infty\frac{h'}{2h}\dd r^*+\int_{-\infty}^\infty\frac{f\mph^2-v}{2h}\dd r^*\rt)\no\\
%&&\red{\text{sech$(-x)=$ sech$(x)$}}\no\\	
    &=&
	\text{sech}^2\lt[\frac{\ln(h)}{2}\bigg|_{-\infty}^\infty+\int_{-\infty}^\infty\frac{f\mph^2-v}{2h}\dd r^*\rt]
    \,\,\,\equiv\,\,\, T_\text{bound2}.\label{Tb2}
\ea
To obtain the above result, we have used the fact that $h'\big(\om^2-v-h^2\big)>0$ in the second line and the triangle inequality: $|x+y|\leq|x|+|y|$ in the third line.
After integrating the argument of the hyperbolic secant function in the last line of Eq.~\eqref{Tb2}, the lower bound of the greybody factor can written as
\ba
    T_\text{bound2}
    =\text{sech}^2\big[X\big],
\ea
where
\ba
    X
    &=&
    \frac{1}{4}\ln\lt(1-\frac{\mph^2}{\om^2}\rt)
    +\frac{(1+q^2)\sqrt{\om^2-\mph^2}-\om(1+3q^2)}{\mph^2q^2}\no\\
    &&+\frac{\Big[\mph^2(1-2q^2-4lq^2-4l^2q^2+q^4)+4 q^2\om^2\Big]}{\mph^3q^3}\no\\
    &&\hspace{1cm}\times\lt[\cot^{-1}\lt\{\frac{q(\om+\sqrt{\om^2-\mph^2})}{\mph}\rt\}-\tan^{-1}\lt\{\frac{\mph(1+q^2)}{2q\sqrt{\om^2-\mph^2}}\rt\}\rt].\label{X}
\ea
Moreover, the integrands $X$ in the Schwarzschild limit ($q\to0$) and the massless limit ($\mph\to0$) can be, respectively, expressed as
\ba
    \lim_{q\to0}X
    &=&
    \frac{1}{4}\ln\lt(1-\frac{\mph^2}{\om^2}\rt)
    -\frac{(1+l+l^2)\om}{\mph^2}
    +\frac{2\om^3+\Big[(3+3l+3l^2)\mph^2-2\om^2\Big]\sqrt{\om^2-\mph^2}}{3\mph^4},\no\\\\
    \lim_{\mph\to0}X
    &=&
    \frac{q^2-3-6l-6l^2}{12\om}.
\ea
Obviously, the bound $T_\text{bound2}$ can be explicitly written in terms of the parameters $q$, $\mph$, and $l$.
It is noticed that the lower bound in Eq.~\eqref{Tb2} should be lower than those in Eqs.~\eqref{Tb1 large m} and $\eqref{Tb1 small m}$ since the inequality properties have been applied for the sake of integrability. 
The behaviors of the bounds of the greybody factor associated with both forms of the function $h$ are shown in Fig.~\ref{fig:Tb}.
%%%%%%%%%%%%%%%%%%%%%%%%%%%%%%
%%%%%%%%%%%%%%%%%%%%%%%%%%%%%%
As illustrated in this figure, the bound associated with $h=\sqrt{\om^2-v}$ is stronger than that associated with $h=\sqrt{\om^2-f\mph^2}$ when $\om$ is sufficiently high.
On the other hand, in the range of low $\om$, only the bound associated with $h=\sqrt{\om^2-f\mph^2}$ can exist.
Hence, they appropriately represent the bound of the greybody factor for different ranges of frequency. 
Specifically, the bound associated with $h=\sqrt{\om^2-f\mph^2}$ is suitable to determine the greybody factor at low frequency. For a sufficiently high frequency, the bound associated with $h=\sqrt{\om^2-v}$ is more suitable, since it provides a closer value to another choice of $h$. 

\begin{figure}[h]
\begin{center}
\includegraphics[scale=0.5]{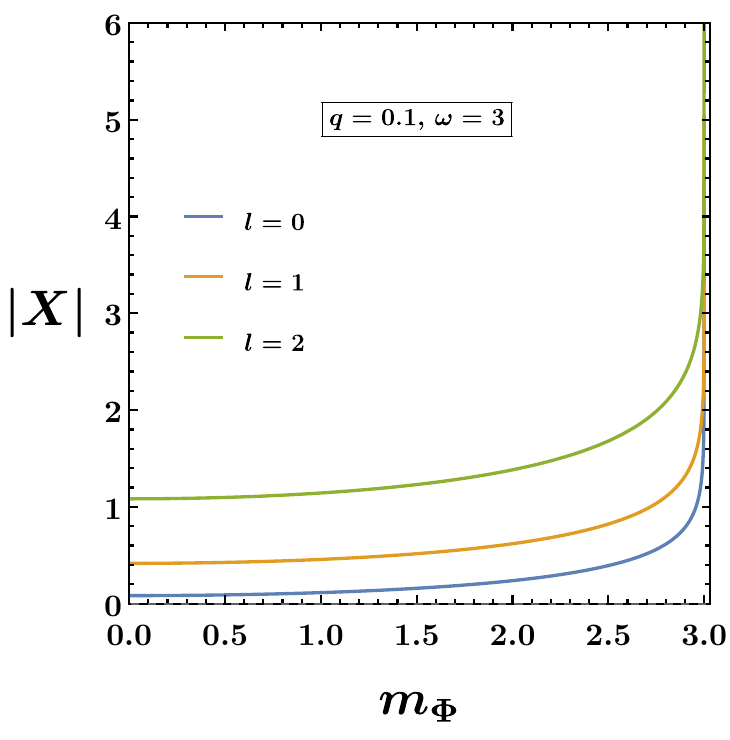}
\end{center}
\vspace{-0.5cm}
\caption{Magnitude of the function $X$ defined in Eq.~\eqref{X} versus $\mph$ for various values of $l$.
}\label{fig: X mph} 
\end{figure}

It is important to note that in both cases of the function $h$, the frequency of the wave is restricted to be greater than the mass of the scalar field, $\omega > m_{\Phi}$. This comes from the fact that the $h$ function must be a positive function by construction. It is worthwhile to note that as seen in Fig.~\ref{fig: X mph}, $X$ is a monotonically increasing function of the scalar field mass. Therefore, due to the behavior of the hyperbolic secant function, the greybody factor always decreases as the field's mass increases. As a result, there is no critical mass which provides the maximum of the greybody factor as found in the literature. In fact, this result is compatible with the analysis in Ref.~\cite{Boonserm:2023oyt} which shows that the existence of the critical mass depends on the existence of the negative part of the potential. In our case, there is no critical mass since the negative part of the potential does not exist.

\begin{figure}[h]
\begin{center}
\includegraphics[scale=0.5]{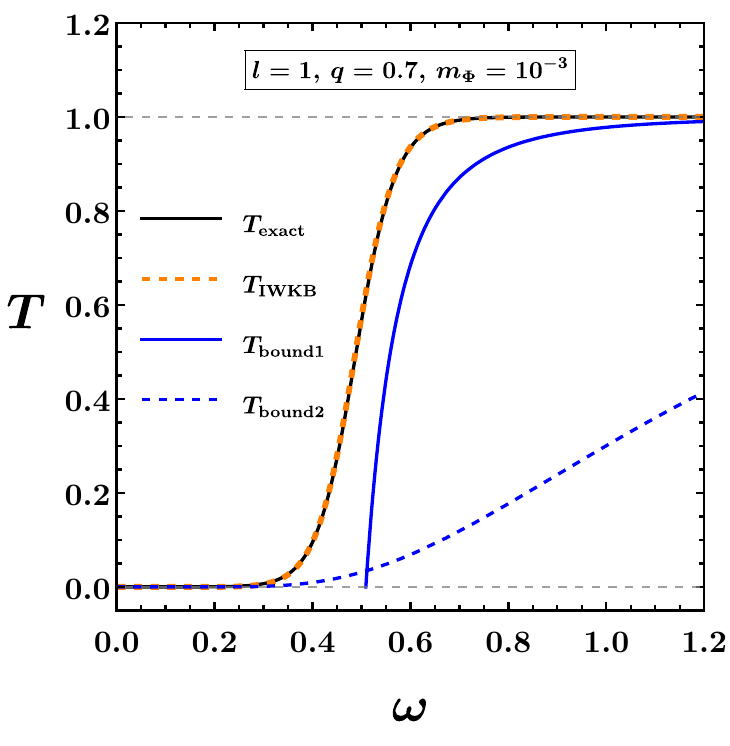}
\end{center}
\vspace{-0.5cm}
\caption{Comparison the results obtained from all methods.
}\label{fig:allT} 
\end{figure}

To check the validity of the rigorous bound method, we compare the bounds with the results approximated by the WKB approximation method. 
In Figs.~\ref{fig:allT}, it is seen that the bound associated with the proposed form of the function $h$ seems to be the proper bound of the greybody factor.
Note also that $T_\text{bound1}$ and $T_\text{bound2}$ are still the proper bound of the greybody factor in the low-frequency regime, i.e., the low frequency WKB approximation is employed.
Unfortunately, the bounds are much lower than the exact value\footnote{
The exact value of the greybody factor can be obtained numerically by matching the wave associated with the scalar perturbed field to the suitable boundary conditions.
The waves at the boundaries are the Taylor's approximation near the event horizon and asymptotically far distance.
Accordingly, we have modified the code that is publicly available in {\tt https://centra.tecnico.ulisboa.pt/network/grit/files/amplification-factors/} by treating the perturbation field as a massive uncharged scalar field.  
}
and the value obtained from the low frequency WKB approximation.
Therefore, the rigorous bound method is useful to deal with the problem analytically.

Unlike the greybody factor obtained from the WKB approximation, the bound proposed in this consideration also provides the results for the $l=0$ mode as seen in the left panels of Fig.~\ref{fig:Tb}. Moreover, one can see that the results from the rigorous bound can be obtained in the high-mass regime (see the top panels of Fig.~\ref{fig:Tb}) while the WKB approximation cannot be applicable. The advantage of the rigorous bound is that it is not only an analytical form, but also can be used for a wider range of parameter values (compared to the WKB approximation method).

%
%%%%%%%%%%%%%%%%%%%%%%%%%%%%%%
%%%%%%%%%%%%%%%%%%%%%%%%%%%%%%

%%%%%%%%%%%%%%%%%%%%%%%%%%%%%%%%%%%%%
\section{Conclusion}
\label{sec:conclu}
%%%%%%%%%%%%%%%%%%%%%%%%%%%%%%%%%%%%%

In this study, we investigate the greybody factors associated with the massive scalar field on the charged black hole spacetime.
It was shown that the dynamics of (the radial part of) the perturbed field can be described by the Regger-Wheeler equation~\eqref{RW eq} which is taken in the Schr\"odinger-like form with the effective potential expressed in Eq.~\eqref{eff pot}.
One then can interpret that the spacetime curvature acts as a barrier where the wave of the perturbed field penetrates through it. 
As a general feature of the potential, it gets higher when $q$ decreases or $\mph$ increases.
Interestingly, the potential can be the monotonically increasing function in $r$ for the massive scenario with sufficiently large $\mph$. 
Unlike the massless scenario, it behaves only as a barrier-like form.

The greybody factor for the corresponding wave function was determined using the WKB approximation and rigorous bound methods.
With the former method, the results shown in Fig.~\ref{fig:Twkb} are reliable because they are in the same trend as the quantum particle does.
The greybody factor decreases as $q$ decreases or $\mph$ increases (causing the higher potential).
Therefore, this method provides a good approximation.
According to this result, it is possible to conclude that the perturbed field with higher mass will encounter a stronger interaction due to spacetime curvature, and then it is more difficult to penetrate through the potential barrier.

For the latter method, we found that the function $h=\sqrt{\om^2-f\mph^2}$ is appropriate to provide the lower bound of the greybody factor in the low-frequency regime while $h=\sqrt{\om^2-v}$ yields a stronger bound in the intermediate-frequency regime.
As we compare these bounds with the results from the WKB approximation method, we have investigated that the results from the rigorous bound method are the proper bounds of the greybody factor.
In addition, the investigation of the bound with $h=\sqrt{\om^2-f\mph^2}$ leads to the result that there is no critical value of $\mph$ in which the greybody factor is maximized.

It is important to remark that, in the large-mass regime as well as the $l=0$ case, the WKB seems to be inapplicable because the potential is not in a barrier-like form (see Ref.~\cite{Konoplya:2019hlu} for further discussion).
However, the rigorous bound method can deal with this situation.
One of the other advantages of the rigorous bound method is that the result (of the bound) is analytic expression, unlike the numerical result in the WKB approximation method.
Therefore, it might be a more powerful tool for studying the physical implications based on black hole perturbation theory.

Further complicated configurations of the test field are also interesting to explore. For instance, for the charged scalar field, it is possible to obtain an interesting phenomenon such as the superradiance.  Moreover, as found in the literature, there are various kinds of black hole solution which are generalizations of the charged black hole. For example, in a dyonic black hole which contains the contribution from the magnetic monopole. It is worthwhile to investigate the greybody factor as well as the superradiance by using the rigorous bound method from the dyonic black hole in order to find a possible way to capture the signature of the magnetic monopole. We leave this investigation for further work.

%We hope that study of the greybody factor of the charged black hole from the massive scalar field may pave the way to a more general criteria to link the physics of black hole to the more fundamental theory or new phenomena of the black hole which may observed in the near future. 

%%%%%%%%%%%%%%%%%%%%%%%%%%%
\section*{Acknowledgement}
We are grateful to Alejandro Saiz Rivera for providing valuable insight and commentary on this work. This research has received funding support from the NSRF via the Program Management Unit for Human Resources \& Institutional Development, Research and Innovation [grant number B39G670016]. SP also acknowledges the support from Sri Trang Thong (STT) scholarship, Faculty of Science, Mahidol University.

%%%%%%%%%%%%%%%%%%%%%%%%%%%%

%%%%%%%%%%%%%%%%%%%%%%%%%%%

\vspace{0.5cm}

\textbf{\textit{Data Availability Statement:}} No Data associated in the manuscript

%%%%%%%%%%%%%%%%%%%%%%%%%%%%

\vspace{0.5cm}

%\bibliographystyle{mybibstyle}

%\bibliographystyle{plain}
%\bibliography{ref_GB_massive}

\end{document}